\documentstyle[prl,aps,array,twocolumn,epsf]{revtex}

\begin{document}
\title{ {\Large {\bf 
The Theory of Superfluidity of $^4$Helium }}}
\author{ J X Zheng-Johansson} 
\address{ H H Wills Physics Laboratory, Tyndall Avenue, 
              Bristol University, Bristol, BS8, 1TL, England }
\date{Submitted to PRL November 12, 1998, Revised December 19, 1998}  
\draft
\bibliographystyle{prsty} 
\def\Bbf{${\bf B}$}
\def\Bbfa{${\bf B}_a$}
\def\Bbfind{${\bf B}_{ind}$}
\def\Ebf{${\bf E}$}
\def\BbfLd{${\bf B_{Ld}}$}
\def\Ebfa{${\bf E}_a$}
\def\Ebfind{${\bf E}_{ind}$}
\def\Jbf{${\bf J}$}
\def\Jbfa{${\bf J}_a$}
\def\Jbfind{${\bf J}_{ind}$}
\def\Jbf{${\bf J}$}
\def\Jbfs{${\bf J_s}$}
\def\kbf{{\bf $ k $}}
\def\kbfsu{{\bf $ k \uparrow $}} 
\def\kbfsd{{\bf $ k \downarrow $}} 
\def\oK{ $ \rm{^\circ}$K} 
\def\rbf{{\bf $ r $}}
\def\xbf{{\bf $ x $}}
\def\alg{{\LARGE a}}
\def\ulg{{\LARGE u}}
\def\elg{{\LARGE e}}
\def\glg{{\LARGE g}}
\def\nlg{{\LARGE n}}
\def\slg{{\LARGE s}}
\def\b[{  {\huge [}  }
\def\Bb{{\bf B}}
\def\Bba{{\bf B}_a}
\def\Bbind{{\bf B}_{ind}}
\def\BbLd{{\bf B_{Ld}}}
\def\Eb{{\bf E}}
\def\Eba{{\bf E}_a}
\def\Ebind{{\bf E}_{ind}}
\def\Jb{{\bf J}}
\def\Jba{{\bf J}_a}
\def\Jbind{{\bf  J}_{ind}}
\def\Jb{{\bf J}}
\def\Jbs{{\bf J_s}}
\def\jbs{{\bf j_s}}
\def\jb{{\bf  j}}
\def\He4{$^4$He}
\def\An{$\mathop{\rm A}\limits^\circ$}
\def\Tc{ $T_c$}
\def\T1{ $T$}
\def\Tlmb{$T_\lambda$}
\def\kb{{\bf  k }}
\def\kbsu{{\bf  k \uparrow }}
\def\kbsd{{\bf  k \downarrow }}
\def\nnr{\nonumber}
\def\ov{\over}
\def\pd{\partial}
\def\rb{{\bf  r }}
\def\Rb{{\bf  R }}
\def\ub{{\bf  u }}
\def\vb{{\bf  v }}
\def\Tlm{T_{\lambda}}
\def\xb{{\bf  x }}
\def\xib{ {\bf  \xi }}
\def\Xib{{\bf  \Xi }}
\def\itS{{\it  S }}
\def\itW{{\it  W }} \def\divd{\bigtriangledown } 
\def\divu{\bigtriangleup} 
\def\divdb{{\bf \bigtriangledown }} \def\divub{{\bf \bigtriangleup }} 
\def\calZ{{\cal Z}}
\maketitle
\begin{abstract}
I present here a  microscopic theory for the superfluidity of $^4$He    
(He II) derived from  experiments,  and  answer its essential questions. 
With a "momenton" model,   the superfluid is shown to feature as a 
"harmonic superfluid".  In which a new bonding type, the  "superfluid bond", is 
formed.  Its  activation causes the anomalous thermal excitation,  the  large 
excess of specific heat, etc. 
The superfluidity mechanism is recognized being connected to a 
quantum confinement effect.  The theory predicts the basic properties of He II 
in overall agreements with experiments. 
The series novel concepts evolved here reverse the current 
perceptions of $^4$He, give significant  impact to the understandings of 
other superfluids.
\end{abstract}
\paragraph*{}
{\it "What is physics, physics is what one observes from  experiments 
\ldots" 
--- By  a condensed matter theorist.} 

$^4$Helium, being small, light and inert, occupies hence a special 
position among the chemical elements. 
The striking manifestation of these extremes in the bulk form is that 
$^4$He does not solidify down to whatever low temperature under 
ambient pressure, transforms  instead at 2.17  \oK \  
from a classical liquid (the normal liquid, or, He I) to the  superfluidity 
phase (the  superfluid, or, He II.)  
The very characteristic property of  the superfluid  is  its absence of  
friction   in some kinematics processes and is also why the name 
"superfluidity" is given.  Enormous amounts of experiments have been made over 
the past 80 years
to learn about the unusual fluid of He II.
To formulate a satisfactory theory for it,  however,  
has presented  as an  intriguing challenge.
The difficulty of the problem is three-fold: He II  is a liquid;  
on top of that it is a quantum liquid;  its ability to flow without dissipation 
is out of normal place among the great majority of condensed  matters. He II 
exemplifies one of the extreme cases that endure the validity of the normal 
physics laws. Briefly speaking, it is generally agreed that there is so far no 
satisfactory 
theory for the  \He4 \ superfluidity. Comprehensive reviews on He II  
are given e.g. in
\cite{wilks}- 
\cite{griffin:1993}.

In this letter, I  present the backbones of the microscopic theory 
of the superfluidity of \He4. \  
The  complementary details and extended contents of the theory 
are presented in  a much elaborate paper \cite{jane:prb}.
The theory has been  formulated from  iterative  contrasts between 
what is  predicted by  an assumed "theory" (solutions of equations of 
motion based on a microscopic model construction) and 
what is observed by the  experiments;  and contrasts between what is 
predicted for one property and its implication to the others;  and 
followed by modifications of the "theory".  
The approach follows treating the superfluid as a "modified solid". 
\\[0.5cm]
\noindent
{\bf  The formulation of the theory based on experimental 
observations} 

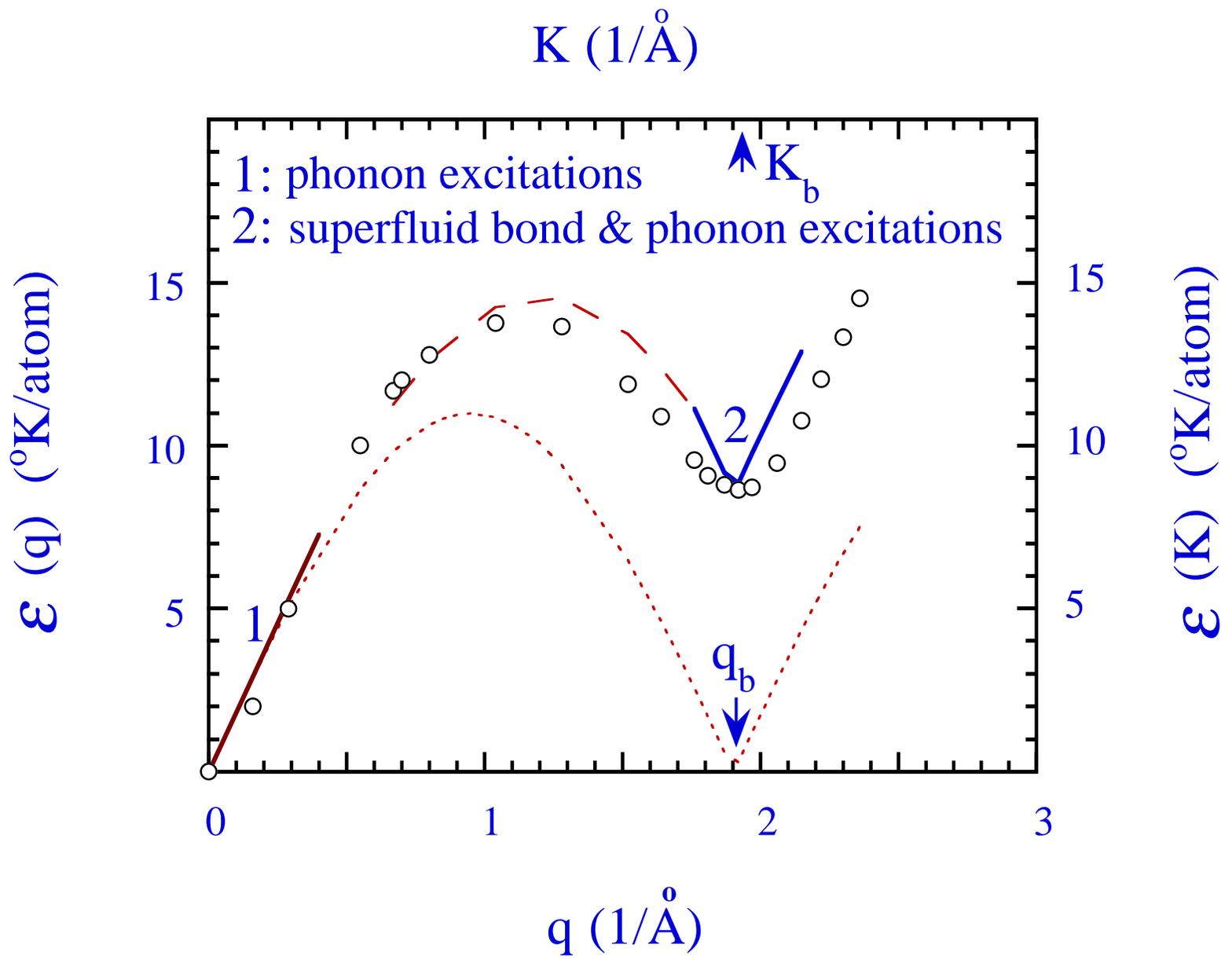
\begin{figure}
\caption{ 
Thermal excitation energy verses phonon momentum $K$ (or neutron 
momentum transfer $q$) for the  superfluid 
of $^4$He,  from  neutron diffraction (circles) at 1.12  $\rm{^oK}$   
and    theory (full lines.)
%
About  full line-1 and the dashed and dotted lines see [5].
}
\label{fig1}
\end{figure}

\label{sec2.1}
\paragraph{ Thermal excitations between $0 \sim 0.6$ \oK, \ phonons}  
\label{sec2.1-i}
In this $T$ region, He II is  observed  in various experiments (see e.g. 
\cite{wilks}) to be virtually a (pure) superfluid. \ In the meantime,  
experiments point to that  He II in this region, hence  the (pure) superfluid, 
predominantly features as assemblies of  harmonic  oscillators about relatively 
fixed sites, assuming some intermediate range of ordering, the associated  
thermal excitations being 
creations of longitudinal single phonons;  I thus term the superfluid  the 
"harmonic superfluid".
 That is consistently pointed to by:
(i)  neutron scattering intensity  for a given momentum transfer, $q$, 
presents \cite{pol:1958} a sharp peak centred at a finite frequency
and the excitation dispersion    resembles \cite{jane:prb} that  of a 
harmonic crystal  except  at $q_b \simeq 1.93$ 1/\An, \ 
see Fig \ref{fig1} (data from \cite{henshaw:w:1961}); \
(ii) the heat capacity has a Debye $T^3$ behaviour \cite{jane:prb} (data 
from \cite{weibes:el});
(iii) a  velocity of  first sound   $ c_1 \simeq 239 $ m/sec  is  
consistently   given \cite{jane:prb}  by    (i) and  (ii) as well as by pulse 
transmission measurements (e.g., \cite{atkins:stasior}.)
Subsequent to  the above, I  infer that  any of the relevant  properties of 
He II  ought to normally result   from the {\it  involvement }  of  the 
single phonon excitations. 

\paragraph{The superfluid  bond}  
\label{sec2.1-ii}

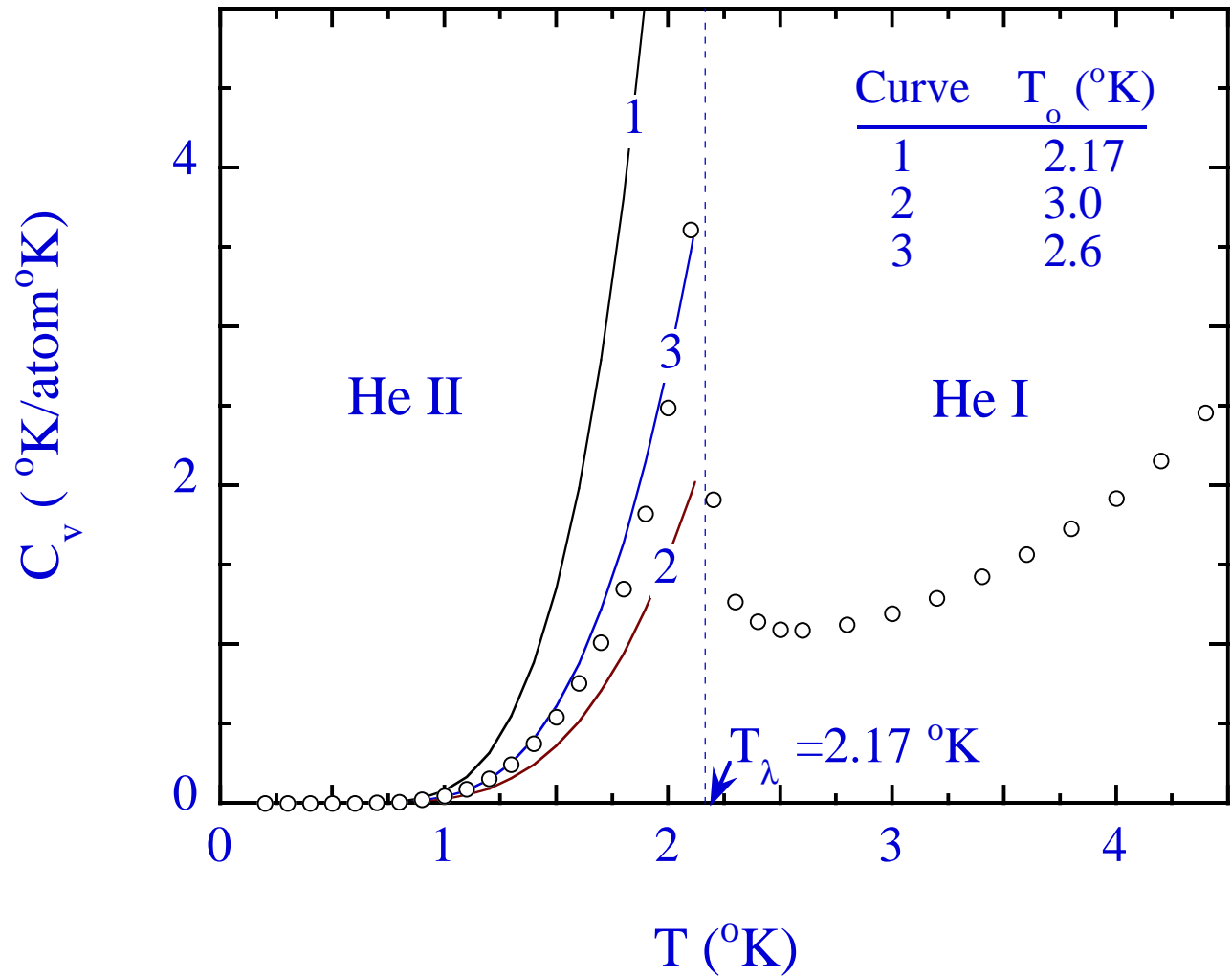
\begin{figure}
\caption{ Specific heat $C_V$ of    superfluid  He II,
from theory (full  lines) and experiments (circles) 
}
\label{fig2}
\end{figure}

The harmonic superfluid character and its associated  suppression of 
atom diffusion into localized vibrations effectively imply that  the 
bonding of He atoms in the superfluidity state has undergone a 
qualitative change and is larger compared to in He I.
I  term the new bonding type as  "superfluid  bond", defined as the 
negative total binding energy per He atom ($U_{os}$), 
and denoted  as $E_s(q) (\simeq - U_{os})$, with $E_b$ ($ 
=E_s(q_b)$) referring to the zero phonon superfluid bond at $q_b, $ cf Fig 1.  
Quantitatively, $E_s \approx E_b$, as well shall see.
Further relevant experimental  indications include:    
(i) a large excess of specific heat of a  $\lambda$ profile presents 
between about 0.6 ($\sim 1$)  to 2.17 \oK, \   see Fig \ref{fig2}  (data from 
\cite{Kramers:1952}, \cite{hill};)
(ii) an anomalous sharp  inelastic neutron scattering peak presents 
\cite{pol:1958}  at $q_b $ of an  energy transfer  of  $\Delta_s \simeq 
8.6$ \oK/atom, \  cf Fig 1; 
(iii) thermodynamic measurement \cite{simon} yields the ground state 
cohesion energy,  $U_{os} \simeq -7.2$ \oK/atom; \ 
(iv) diffusion  of a He ion in the superfluid \cite{Reif:el:1960} involves  
an activation energy of $ 8.1 \sim 8.6$ \oK/atom.
(i)-(iv) consistently point to the presence of a superfluid bond,   
$E_b  $ ($\simeq  - U_{os})  =   \Delta_s  - E_v   \simeq  7.2$   
\oK/atom, \ $E_v$ being a small energy needed for site creation.
It follows that, between $0.6$ ($\sim 1$) $\sim 2.17 $ \oK, He II is a 
co-existent of the superfluid and the normal fluid, 
and a large excess  heat is required for heating up and is consumed 
mainly to convert the super- to the normal fluid  via activation of the  
superfluid bond.

\paragraph{The momenton wave and  collective phonons}   
\label{sec2.1-iii}

Despite of  {\it a} and {\it b},   He II  is a liquid. To reflect  this 
solid-liquid duality  but with   {\it solid }  being much emphasised,
I introduce a "momenton" description of the superfluid.
A "momenton" is defined as   an  aggregation of a macroscopically 
large number of  He  atoms   which  rarely  diffuse individually but primarily 
oscillate about sites fixed on  the centre of mass coordinates of their 
momenton.  
Superposed to  the  {\it independent }  oscillations (the single phonon 
events) are  slow {\it in-phase} and {\it simultaneous}  vibrations of all atoms 
following  the momenton,  which   give rise to collective phonons 
whose energy contribution though is usually negligible \cite{jane:prb}.
The superfluid atoms also translate  in  such a collective fashion.   At 
$T$ (0.6 $\sim$ 2.17 \oK) \ where the two fluids co-exist,  they  each  
aggregates in large islands of regions, rather than  intermingles on an atomic 
scale. 
Relevant experimental  indications include:  a slow (second) sound wave 
can propagate in He II  and its pulse broadens as $T$ falls (e.g. 
\cite{peshkov});  
 thermal  conductivity in He II  is exceptionally large,  and not 
defined by  temperature gradient.
\label{sec2.3}
\\[0.5cm]
\noindent
{\bf The equations of motion } 

\noindent
Based on b and c, I can readily simplify the complex superfluid atom 
motion  by  coordinates decomposition,  and  in the semi-classical forms  
we have: 
\begin{eqnarray}
\label{eq7}
&       &
-\alpha_1 \sum_{j-1,j,j+1} \ub_i =  m {\pd ^2 \ub_j \over  \pd t^2} 
\\ 
\label{eq8}
&       &
D \divd^2 n(\Rb) - {\partial n(\Rb) \over \partial t} =0   
\\ 
\label{eq9}
& & -\alpha_2 \sum_{j-1,j,j+1} \xib_i =  M_j {\pd ^2 \xib_j \over  \pd 
t^2},
\quad \quad j = 1, 2, \ldots, N_M
\\
\label{eq10}
  & &   F_{12} - \Upsilon 
                  = \sum_i^{N_M} M_i {\pd ^2 \Xib \over  \pd t^2} 
\end{eqnarray}

\noindent
Their quantum mechanical representations  can be  similarly 
decomposed \cite{jane:prb}.
In above,   the He  atom  absolute position $\rb$  is a vector sum  of  
four relative  coordinates:
$
       \rb (x,y,z)  = \ub + \Rb  + \xib + \Xib,  
$
as schematically illustrated in Fig  3. 
Of Eqs (\ref{eq7})-(\ref{eq10}),  (\ref{eq7}) and (\ref{eq9})  describe  
the vibrations of a He atom and a momenton, for their solutions and 
discussions see \cite{jane:prb}. (\ref{eq8}) and (\ref{eq10}) are discussed 
below.

\noindent
\paragraph*{{\it a.}  Excitation of superfluid bond and the total  
excitation }
\label{sec2.4-2}
Under thermal equilibrium, Eq (\ref{eq8}) describes, in terms of the 
liquid  density $n(\Rb)$,   a single atom translation, $\Rb$, relative 
to the momenton coordinates, $\Xib$. 
It can be  thermal diffusion, or atom displacement  caused by an external 
perturbation such as an incident neutron, 
or an applied  field, etc. These all involve a  probability  to excite  an 
atom from a  bound state of the bonding energy $E_b$,  amounting to the 
proportion to  the Boltzmann factor,
\begin{equation}  
\label{eq2-10}
P_s(\Rb \rightarrow \Rb')   \propto  D \propto e^{-{ \Delta_s \ov k_B 
T }} 
= e^{-{ (E_b + E_v) \ov k_B T }}
\end{equation}
Due to certain lattice ordering in the  superfluid, the superfluid  bond 
excitation is dependent on the   phonon wavevector,  $K$, as  being 
explicitly shown in  the neutron data.  And the maximum  intensity 
occurs at $K_b \propto 2\pi/a \simeq 1.93 $ 1/\An, \ i.e.,  the reciprocal 
lattice vector (expressed for  a cubic lattice.)  If it were a solid (of bonding 
energy: $\sim 10^5$ \oK/atom), it would cause  neutron  
an elastic  Bragg  scattering at $K_b$.  Due to an intermediate  bond 
strength ($\sim 8.6$ \oK/atom)  which falls below the thermal neutron 
energy (60 to 1160 \oK/atom), \  the superfluid 
instead scatters  a  neutron   at $K_b$ with  an  energy transfer of 
$\Delta_s = (E_b + E_v)  \simeq 8.6$ \oK/atom, \ see Fig 1. 
For  $ |K-K_b| > 0$, a neutron begins to  excite phonons more   
readily, thus $ P_s$ drops   rapidly   within a  narrow  width $\sim \sigma$.  
So that, for $ |K-K_b| > \sigma$,   $P_s \simeq 0$ and  the full excitation 
involves primarily  phonons: ${\cal E}(K) \approx \hbar \omega_{ph}$; for $ |K-
K_b| <\sigma$, 
\begin{eqnarray}
\label{eqex}
{\cal E} (K) & & = \Delta_s + \hbar \omega_{ph}
            \approx     E_b(K_b)  + E_v + {\hbar c_1 \ov k_B} |K-K_b|     
 \\ \nnr              
& &      \approx  8.6 +  18.2  |K-1.93| \quad\quad  
(\rm{^oK/atom}) 
 \end{eqnarray}
 where  $\hbar = h /2\pi$ and  $k_B$ being the Planck's and  Boltzmann 
constants;  $\hbar \omega_{ph} $ being expressed for  the acoustic  
phonons.
The parameterized Eq (\ref{eqex}),  obtained using values of $\Delta_s 
$ and $c_1$   given earlier,   is as plotted  in Fig 1  
in the vicinity of  $K_b $ 
(full line-2), and is seen to reproduce the experiment curve here 
reasonably well.

\paragraph*{ b Superfluid flow,  translation of  momentons and  atoms 
} 
 For a  uniform and steady   flow, Eq (\ref{eq10}) describes  the 
translation of a momenton or an atom or the flow under  an 
external  pressure difference, $F_{12}$,   $\Upsilon $ being the viscous  
resistance, $\sum_i^{N_M} M_i = M_s$  the flow mass, $M_i$ the 
mass of the $\imath$th of the total $N_M$ momentons.
For the case as above,  there is, $<\rb> \ = \Xib $ \   ($< >$ refers to time 
average over an infinitesimal $\Xib$ translation.)
Provided   the translation  velocity,  $v_s$,  
is below a critical value,  so that $\Upsilon \equiv  0$,   the  solution of  Eq 
(\ref{eq10}) describes the non-dissipative superfluid translation at the absence 
of $F_{12}$: 
\begin{eqnarray}  
 v_s = {\pd  <\rb> \ \over  \pd t} 
          \equiv  \rm{constant}
\end{eqnarray}
The quantum mechanical counterpart describing a He atom translation  
is an {\it effective}  planewave \cite{jane:prb},
\begin{eqnarray} 
\label{eq2-13c}
\Psi(\rb) = C e^{\imath({\bf k_s} <\rb> - \omega t)}
\end{eqnarray} 
where  $k_s  = m v_s / \hbar $, $m$ is He atomic mass. $k_s$ and 
$v_s$ are 
both {\it effective} in a sense that an atom is accelerated with an 
inertia equal to the  flow mass,  $M_s$

\noindent
\label{sec2.6}
\\[0.5cm]
\noindent
{\bf The mechanism  of  superfluidity } 

\noindent
In the harmonic superfluid, conveyed in the (longitudinal) elastic wave
propagation is that of phonons. Normally phonon excitation 
energy: $\Delta \omega(K_{ph}) \simeq c_1 \Delta K_{ph}$ can be 
arbitrarily small for  $\Delta  K_{ph}$ being small.
%
%
When in a narrow channel, however, its width $d$ then sets a maximum 
wavelength $\lambda_{max}$ for phonons able to propagate in the 
superfluid, which corresponding to  the minimum  phonon wavevector 
and an associated  energy gap,
\begin{eqnarray}
\label{eq3-10}
(K_{ph})_{min} = {2 \pi \ov  \lambda_{max} }=  {2 \pi  \ov d}, 
\quad     \rm{and}   \quad
\Delta_{ph} =\hbar \omega_{ph}((K_{ph})_{min})  
\end{eqnarray}
this being referred as one type of the "quantum confinement effects" 
(similar effect for free particle like atoms is treated in \cite{jane:prb}.) 
It leads to that no lower modes phonons than $(K_{ph})_{min}$  can 
be excited in the confined superfluid sample 
here by the flow and wall interaction 
{\it This is the cause for  superfluidity motion of He II} (at larger $d$.)  
This  mechanism  is pointed to by the two characteristic  
experimental demonstrations:
(i)  a He II rotation current  persists over the entire experiment 
duration (12 hrs) \cite{JD:Reppy}.
(ii) He II  translation  exhibits non-dissipative superfluidity only  in 
narrow channels and below a critical velocity, $v_c$, and $v_c(d) $  increases 
as  $d$ falls, see Fig 4  (data from \cite{wilks}).

I now establish the first principles  criteria for it and evaluate $v_c(d)$.
The energy exchange between  the flow and the wall is via  $N_{surf}$ 
atoms at the interface,  each having the flow inertia $M_s$; and hence is via an 
effective mass:
$ m_{eff} = M_s/N_{surf}  =   m d  / 4 a.  $ 
Upon  collisions with the wall,   the flow  loses its 
translation momentum (initially $k_s, v_s$) and energy ($\propto M_s 
v_s^2/2$),  so that  in the severest case being stopped 
($k_s'=v_s'=0$.)
Being  inert, the superfluid atoms   
can be approximated as not interacting with the wall except for the moments of 
collisions. 
The heat thus produced is converted to disordered  atom vibrations in 
the bulk --- the creation of  phonons,  of the energy  $\hbar c_1 K_{ph}$ in the 
case of low excitations. 
A non-dissipative  superfluid translation hence demands the  two 
simultaneous conditions:
\begin{eqnarray}  
\label{eq3-8}
& k_{s} \le \left(K_{ph}\right)_{min},  \\
\label{eq3-8b}
& m_{eff} v_{s}^2 /2   \le \Delta_{ph} = \left( {\hbar c_1 
K_{ph}}\right)_{min} 
\end{eqnarray}  
Using  (\ref{eq3-10})  in the more restrictive Eq  (\ref{eq3-8b})  
yields the critical velocity:
\begin{eqnarray}  
\label{eq3-9}
 v_{c} 
            = \sqrt{{8 h c_1 a \ov m}}  \quad {1 \ov d}
            =  2.62 \times 10^{-7}  {1  \ov  d \ (\rm{m})  }   \quad \quad  
(\rm{m/sec})
\end{eqnarray} 

\noindent 
 The parameterized  Eq (\ref{eq3-9}) is  obtained with:
$ c_1 = 239$   m/sec,   and  $a = 3.6$ \An \ (the He II atomic spacing.)
The  predicted  $v_c(d)$, full line in Fig 4,  is seen to agree well 
with experiments  for larger $d$.
For low $d$, improved $v_c(d)$  (dashed line) has been obtained  
\cite{jane:prb} by   treating the He atoms appropriately as free 
particles. 
\\[0.5cm]
Based on the microscopic theory outlined above, a statistical 
thermodynamic description of He II is also given in
\cite{jane:prb}. Amongst others it predicts the experimental properties, 
$T_{\lambda}$ and $C_V$ (full lines of Fig 1), etc, with 
overall satisfaction. Some of the results were used in the 
discussions here, where Ref \cite{jane:prb} is referred. 
\\[0.5cm]
\noindent
I end this letter with some complementary remarks. 
(1) The "harmonic superfluid" character  implies that the (pure) 
superfluid is thermally excited to a full extent in the longitudinally modes, no 
less than  any solid. So, (1.a) it is not an assembly of  
{\it motion-less} atoms that was thought to lead to superfluidity, (1.b) 
neither its {\it reminiscing   a harmonic solid alone } explains 
superfluidity since a solid does not exhibit  "superfluidity".
(2) It follows, from the  substantial atomic bonding,  that the superfluid 
internal viscosity is large, especially  compared to that of He I.   
So the superfluid is not an assembly of {\it non-interacting atoms that 
would result in zero internal viscosity}. Particularly, Stoke's theorem hence 
does not apply to the superfluid and to ensure   non-turbulence 
thus does not require a {\it total circulation zero}, which being instead 
quantized \cite{jane:prb}.
 It follows that  no {\it local vortices} present in the superfluid. 
(3) Based on quantum mechanical solutions, it can be 
elucidated \cite{jane:prb} that the "superfluid bond"  results  from the  
quantum He atom mediation  among neighbouring sites, and it differs 
from the  known chemical bonds that are mediated via electrons.
%
%
(4) What eases one to  face  the observed bulk properties   
is  the recognition of the superfluidity mechanism here,  
which appears made conclusive by the good agreement 
between the first principles evaluation of $v_c(d)$ (without any 
adjustable parameter) and experiments. 
(5) The same as the bulk phenomena are shown in the theory
to be  interrelated,  quantitative  evaluations of the He II  properties    
mainly are based on two  common parameters, $c_1$ or/and $\Delta_s$  
which being well determined from  experiments, and are fully or 
partially satisfactory.
(6) The  essential properties of  the superfluid of  $^4$He 
have been explained based on a unified theory ground; and  the  physics 
necessitating  the depiction of the 
extreme case of $^4$He  is shown to  retain   the physics laws that apply 
to normal matters.
%
%
\\[0.5cm]
\noindent
The author  is supported by the Fellowships from
the Swedish Natural Science Research Council and the Wallenbergs 
Stiftelse. 
My theory   development  has been routinely communicated to 
B Johansson, P-I Johansson
and V Heine, throughout stimulation   has been given by them.
Supports   have been given by    M Springford,   P Meeson,    
B Gyorffy, J A Wilson,  K Sk\"old,  and O Eriksson.
A conversation with P Meeson  has  caught  my attention  to the  
persistent current  experiment. 


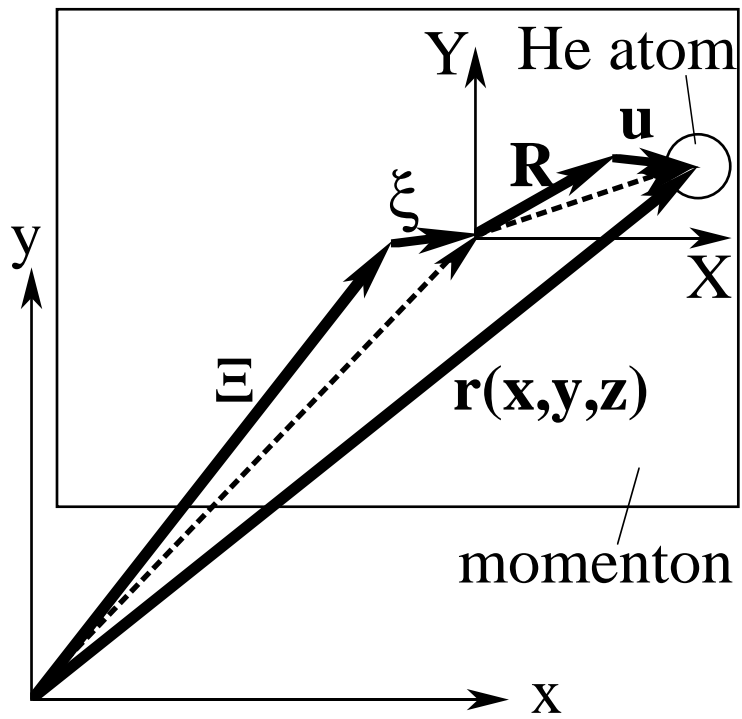
\begin{figure}
\caption{  Illustration of   relations of the relative coordinates $\ub$, 
$\xib$, $\Xib $ and the absolute position ${\bf r}$ of a superfluid atom. }
\label{fig3}
\end{figure}

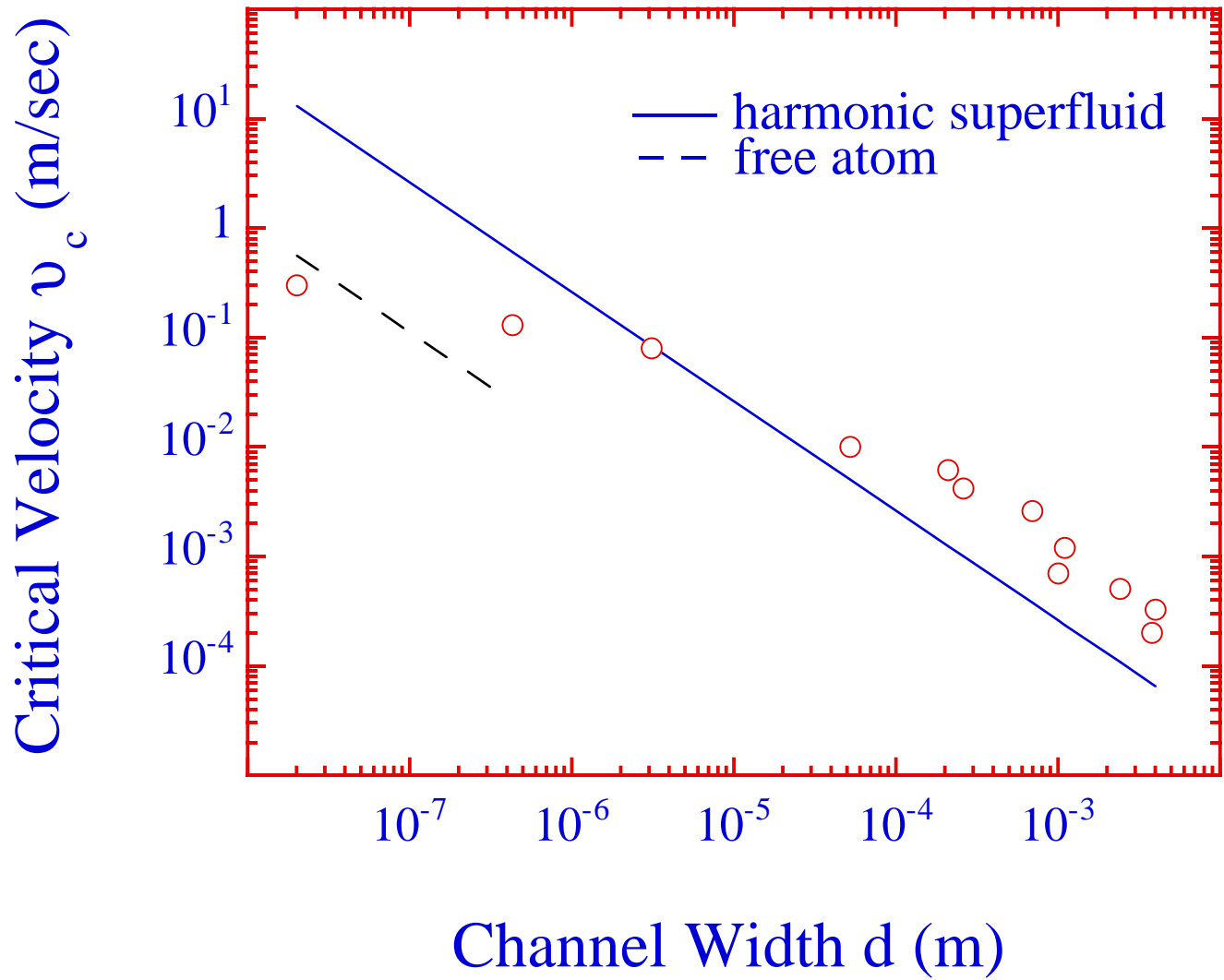
\begin{figure}
\caption
{
Critical velocity $v_c (d) $  
for the  superfluid of $^4$He,   from theory  (full and dashed lines)
and experiments (circles) at about 1.4 $\rm{^oK}$. 
}
\label{fig4}
\end{figure}
\end{document}